%% file: acl2020.tex
\documentclass[11pt,a4paper]{article}
\usepackage[hyperref]{acl2020}
\usepackage{times}
\usepackage{latexsym}
\usepackage{enumitem}
\usepackage{paralist}

\usepackage{url}
\usepackage{comment}
\usepackage[flushleft]{threeparttable}
\usepackage{graphicx}
\usepackage{amsmath}
\usepackage{subfig}
\usepackage{tabularx}
\usepackage{mathrsfs}
\usepackage{multirow}
\usepackage{algorithm}
\usepackage{enumitem}
\usepackage{xcolor}
\usepackage{listings}
\usepackage{hyperref}
\usepackage[frozencache,cachedir=.]{minted}

\usepackage{amssymb}
\usepackage{pifont}
\newcommand{\cmark}{\ding{51}}%
\newcommand{\xmark}{\ding{55}}%

\usepackage[utf8]{inputenc}
\usepackage{cleveref}
\crefname{section}{§}{§§}
\Crefname{section}{§}{§§}

\usepackage{xcolor} 

\definecolor{MyColor}{RGB}{50, 100, 250}
\definecolor{Orange}{RGB}{244, 101, 66}
\definecolor{Red}{RGB}{255, 0, 0}
\definecolor{Green}{RGB}{0, 255, 0}
\definecolor{Blue}{RGB}{0, 0, 255}

\title{A Transformer-based Approach for Source Code Summarization}

\aclfinalcopy 

\author{
Wasi Uddin Ahmad \\
University of California, Los Angeles \\
\texttt{wasiahmad@cs.ucla.edu} \\
\And
Saikat Chakraborty \\
Columbia University \\
\texttt{saikatc@cs.columbia.edu} \\
\AND 
Baishakhi Ray \\
Columbia University \\
\texttt{rayb@cs.columbia.edu} \\
\And
Kai-Wei Chang \\
University of California, Los Angeles \\
\texttt{kwchang@cs.ucla.edu} \\
}

\begin{document}

\setlength{\abovedisplayskip}{5pt}
\setlength{\belowdisplayskip}{5pt}

\maketitle

\begin{abstract}
Generating a readable summary that describes the functionality of a program is known as source code summarization. In this task, learning code representation by modeling the pairwise relationship between code tokens to capture their long-range dependencies is crucial. To learn code representation for summarization, we explore the Transformer model that uses a self-attention mechanism and has shown to be effective in capturing long-range dependencies. In this work, we show that despite the approach is simple, it outperforms the state-of-the-art techniques by a significant margin. We perform extensive analysis and ablation studies that reveal several important findings, e.g., the absolute encoding of source code tokens' position hinders, while relative encoding significantly improves the summarization performance. We have made our code publicly available\footnote{https://github.com/wasiahmad/NeuralCodeSum} to facilitate future research.


\end{abstract} 

\input{introduction}
\input{method}

\input{experiment}

\input{relwork}

\input{conclusion}

\section*{Acknowledgments}
This work was supported in part by National Science Foundation Grant OAC 1920462, CCF 1845893, CCF 1822965, CNS 1842456.

\bibliography{acl2020}
\bibliographystyle{acl_natbib}

\cleardoublepage
\input{appendix}

\end{document}

%% file: introduction.tex
\section{Introduction}
Program comprehension is an indispensable ingredient of software development and maintenance \cite{xia2017measuring}.
A natural language summary of source code facilitates program comprehension by reducing developers' efforts significantly \cite{sridhara2010towards}.
Source code summarization refers to the task of creating readable summaries that describe the functionality of a program. 

With the advancement of deep learning and the availability of large-scale data through a vast number of open-source repositories, automatic source code summarizing has drawn attention from researchers.
Most of the neural approaches generate source code summaries in a sequence-to-sequence fashion.
One of the initial works~\citet{iyer2016summarizing} trained an embedding matrix to represent the individual code tokens and combine them with a Recurrent Neural Network (RNN) via an attention mechanism to generate a natural language summary.
Subsequent works \cite{liang2018automatic, hu2018deep, hu2018summarizing} adopted the traditional RNN-based sequence-to-sequence network \cite{sutskever2014sequence} with attention mechanism \cite{luong2015effective} on different abstractions of code. 

The RNN-based sequence models have two limitations in learning source code representations. 
First, they do not model the non-sequential structure of source code as they process the code tokens sequentially.
Second, source code can be very long, and thus RNN-based models may fail to capture the long-range dependencies between code tokens.
In contrast to the RNN-based models, \emph{Transformer} \cite{vaswani2017attention}, which leverages 
self-attention mechanism, can capture long-range dependencies. 
Transformers have been shown to perform well on many natural language generation tasks such as machine translation \cite{wang2019learning}, text summarization \cite{you-etal-2019-improving}, story generation \cite{fan2018hierarchical}, etc.

To learn the order of tokens in a sequence or to model the relationship between tokens,
Transformer requires to be injected with positional encodings \cite{vaswani2017attention, shaw2018self, shiv2019novel}.
In this work, we show that, by modeling the pairwise relationship between source code tokens using relative position representation \cite{shaw2018self}, we can achieve significant improvements over learning sequence information of code tokens using absolute position representation \cite{vaswani2017attention}.



We want to emphasize that our proposed approach is simple but effective as it outperforms the fancy and sophisticated state-of-the-art source code summarization techniques by a significant margin.
We perform experiments on two well-studied datasets collected from GitHub, and the results endorse the effectiveness of our approach over the state-of-the-art solutions.
In addition, we provide a detailed ablation study to quantify the effect of several design choices in the Transformer to deliver a strong baseline for future research.

%% file: method.tex
\section{Proposed Approach}
We propose to use \emph{Transformer} \cite{vaswani2017attention} to generate a natural language summary given a piece of source code.
Both the code and summary is a sequence of tokens that are represented by a sequence of vectors, $\mathbf{x}=(x_1, \dots, x_n)$ where $x_i \in R^{d_{model}}$.
In this section, we briefly describe the Transformer architecture (\cref{sec:arch}) and how to model the order of source code tokens or their pairwise relationship (\cref{sec:pos_rep}) in Transformer. 

\subsection{Architecture}
\label{sec:arch}
The Transformer consists of stacked multi-head attention and parameterized linear transformation layers for both the encoder and decoder.
At each layer, the multi-head attention employs $h$ attention heads and performs the self-attention mechanism.

\smallskip\noindent
\textbf{Self-Attention.}
We describe the self-attention mechanism based on \citet{shaw2018self}. 
In each attention head, the sequence of input vectors, $\mathbf{x}=(x_1, \dots, x_n)$ where $x_i \in R^{d_{model}}$ are transformed into the sequence of output vectors, $\mathbf{o}=(o_1, \dots, o_n)$ where $o_i \in R^{d_k}$ as:
\begin{align*}
o_i &= \sum^{n}_{j=1} \alpha_{ij} (x_j W^V), \\
e_{ij} &= \frac{x_i W^Q (x_j W^K)^T}{\sqrt{d_k}},
\end{align*}
where $\alpha_{ij} = \frac{\exp e_{ij}}{\sum^{n}_{k=1} \exp e_{ik}}$ and $W^Q, W^K \in R^{d_{model} \times d_k}, W^V \in R^{d_{model} \times d_v}$
are the parameters that are unique per layer and attention head.


\smallskip\noindent
\textbf{Copy Attention.}
We incorporate the copying mechanism \cite{see2017get} in the Transformer to allow both generating words from vocabulary and copying from the input source code.
We use an additional attention layer to learn the copy distribution on top of the decoder stack \cite{nishida2019multi}.
The copy attention enables the Transformer to copy rare tokens (e.g., function names, variable names) from source code and thus improves the summarization performance significantly (\cref{subsec:results}).

\input{table/table1.tex}

\subsection{Position Representations}
\label{sec:pos_rep}
Now, we discuss how to learn the order of source code tokens or model their pairwise relationship.

\smallskip\noindent
\textbf{Encoding absolute position.}
To allow the Transformer to utilize the order information of source code tokens, we train an embedding matrix $W^{P_e}$ that learns to encode tokens' absolute positions into vectors of dimension $d_{model}$.
However, we show that capturing the order of code tokens is not helpful to learn source code representations and leads to poor summarization performance (\cref{subsec:impact_of_pos_rep}).

It is important to note that we train another embedding matrix $W^{P_d}$ that learns to encode the absolute positions of summary tokens.\footnote{In this work, we do not study alternative ways of learning position representation for the summary tokens.}

\input{table/table2}

\smallskip\noindent
\textbf{Encoding pairwise relationship.}
\label{subsec:rel_pos_desc}
The semantic representation of a code does not rely on the absolute positions of its tokens. Instead, their mutual interactions influence the meaning of the source code. 
For instance, semantic meaning of the expressions {\tt a+b} and {\tt b+a} are the same. 


To encode the pairwise relationships between input elements, \citet{shaw2018self} extended the self-attention mechanism as follows.
\begin{align*}
o_i &= \sum^{n}_{j=1} \alpha_{ij} (x_j W^V + a_{ij}^V),\\
e_{ij} &= \frac{x_i W^Q (x_j W^K + a_{ij}^K)^T}{\sqrt{d_k}},
\end{align*}
where, $a_{ij}^V$ and $a_{ij}^K$ are relative positional representations for the two position $i$ and $j$.
\citet{shaw2018self} suggested clipping the maximum relative position to a maximum absolute value of $k$ as they hypothesize that precise relative position information is not useful beyond a certain distance.
\begin{equation*}
a_{ij}^K = w_{clip(j-i, k)}^K, a_{ij}^V = w_{clip(j-i, k)}^V,
\end{equation*}
\begin{equation*}
clip(x, k) = \max(-k, \min(k, x)).
\end{equation*}

Hence, we learn $2k+1$ relative position representations: $(w_{-k}^K, \dots, w_k^K)$, and $(w_{-k}^V, \dots, w_k^V)$.

In this work, we study an alternative of the relative position representations that ignores the \emph{directional} information \cite{ahmad2019cross}.
In other words, the information whether the $j$'th token is on the left or right of the $i$'th token is ignored.
\begin{equation*}
a_{ij}^K = w_{clip(|j-i|, k)}^K, a_{ij}^V = w_{clip(|j-i|, k)}^V,
\end{equation*}
\begin{equation*}
clip(x, k) = \min(|x|, k).
\end{equation*}

%% file: table/table1.tex

\begin{table}[t]
\centering
\resizebox{\linewidth}{!}{%
\small
\begin{tabular}{l|c|c}
\hline
Dataset & Java & Python \\ 
\hline
Train &  69,708 & 55,538 \\
Validation & 8,714 & 18,505 \\
Test & 8,714 & 18,502 \\ \hline
Unique tokens in code & 66,650 & 307,596 \\
Unique tokens in summary &  46,895 & 56,189 \\ 
\hline
Avg. tokens in code & 120.16 & 47.98  \\
Avg. tokens in summary &  17.73 & 9.48 \\
\hline
\end{tabular}
}
\caption{Statistics of the experiment datasets. We thank the authors of \citet{wei2019code} for kindly sharing the Python dataset splits.
The Java dataset splits are publicly available.
}
\label{table:statistics}
\vspace{-2mm}
\end{table}

%% file: table/table2.tex
\begin{table*}[!ht]
\centering
\resizebox{\linewidth}{!}{%
\small
\begin{tabular}{l|c@{\hskip 0.125in} c@{\hskip 0.125in} c|c@{\hskip 0.125in} c@{\hskip 0.125in} c}
\hline
\multirow{ 2}{*}{Methods} & \multicolumn{3}{c|}{Java} & \multicolumn{3}{c}{Python} \\ 
\cline{2-7}
& BLEU & METEOR & ROUGE-L & BLEU & METEOR & ROUGE-L \\ 
\hline
CODE-NN \cite{iyer2016summarizing} &  27.60 & 12.61 & 41.10 & 17.36 & 09.29 & 37.81 \\
Tree2Seq \cite{eriguchi2016tree} & 37.88 & 22.55 & 51.50 & 20.07 & 08.96 & 35.64 \\
RL+Hybrid2Seq \cite{wan2018improving} & 38.22 & 22.75 & 51.91 & 19.28 & 09.75 & 39.34 \\
DeepCom \cite{hu2018deep} & 39.75 & 23.06 &  52.67 & 20.78 & 09.98 & 37.35 \\
API+CODE \cite{hu2018summarizing} & 41.31 & 23.73 & 52.25 & 15.36 & 08.57 & 33.65 \\
Dual Model \cite{wei2019code} & 42.39 & 25.77 & 53.61 & 21.80 & 11.14 & 39.45 \\
\hline
\multicolumn{7}{l}{Our models and ablation study} \\
\hline
Base Model & 43.41 & 25.91 & 52.71 & 31.08 & 18.57 & 44.31 \\
Full Model & {\bf 44.58} & {\bf 26.43} & {\bf 54.76} & {\bf 32.52} & {\bf 19.77} & {\bf 46.73} \\
\hline
Full Model w/o  Relative Position & 44.26 & 26.23 & 53.58 & 31.38 & 18.69 & 44.68 \\
Full Model w/o Copy Attention & 44.14 & 26.34 & 53.95 & 31.64 & 19.17 & 45.42 \\
\hline
\end{tabular}
}
\caption{Comparison of our proposed approach with the baseline methods. The results of the baseline methods are directly reported from \cite{wei2019code}.
The ``Base Model'' refers to the vanilla Transformer (uses absolute position representations) and the ``Full Model'' uses relative position representations and includes copy attention.}
\label{table:results}
\end{table*}

%% file: experiment.tex
\section{Experiment}

\subsection{Setup}

\smallskip\noindent
\textbf{Datasets and Pre-processing.}
We conduct our experiments on a Java dataset \cite{hu2018summarizing} and a Python dataset \cite{wan2018improving}. 
The statistics of the two datasets are shown in Table \ref{table:statistics}.
In addition to the pre-processing steps followed by \citet{wei2019code}, 
we split source code tokens of the form \emph{CamelCase} and \emph{snake\_case} to respective sub-tokens\footnote{The \emph{CamelCase} and \emph{snake\_case} tokenization reduces the vocabulary significantly. For example, the number of unique tokens in Java source code reduced from 292,626 to 66,650.}.
We show that such a split of code tokens improves the summarization performance.

\smallskip\noindent
\textbf{Metrics.}
We evaluate the source code summarization performance using three metrics, BLEU \cite{papineni2002bleu}, METEOR \cite{banerjee2005meteor}, and ROUGE-L \cite{lin2004rouge}.

\smallskip\noindent
\textbf{Baselines.}
We compare our Transformer-based source code summarization approach with five baseline methods reported in \citet{wei2019code} and their proposed Dual model. We refer the readers to \cite{wei2019code} for the details about the hyperparameter of all the baseline methods.

\smallskip\noindent
\textbf{Hyper-parameters.}
We follow \citet{wei2019code} to set the maximum lengths and vocabulary sizes for code and summaries in both the datasets.
We train the Transformer models using Adam optimizer \cite{kingma2014adam} with an initial learning rate of $10^{-4}$.
We set the mini-batch size and dropout rate to 32 and 0.2, respectively.
We train the Transformer models for a maximum of 200 epochs and perform early stop if the validation performance does not improve for 20 consecutive iterations.
We use a beam search during inference and set the beam size to 4.
Detailed hyper-parameter settings can be found in Appendix A.

\subsection{Results and Analysis}
\label{subsec:results}

\smallskip\noindent
\textbf{Overall results.}
The overall results of our proposed model and baselines are presented in Table \ref{table:results}. The result shows that the \emph{Base} model outperforms the baselines (except for ROUGE-L in java), while the \emph{Full} model improves the performance further.\footnote{We observe a more significant gain on the Python dataset and a detailed discussion on it is provided in Appendix \ref{app:rnn_vs_trans}.}
We ran the Base model on the original datasets (without splitting the \emph{CamelCase} and \emph{snake\_case} code tokens) and observed that the performance drops by 0.60, 0.72 BLEU and 1.66, 2.09 ROUGE-L points for the Java and Python datasets respectively.
We provide a few qualitative examples in Appendix \ref{app:qual_exam} showing the usefulness of the Full model over the Base model.

Unlike the baseline approaches, our proposed model 
employs the copy attention mechanism.
As shown in Table \ref{table:results}, the copy attention improves the performance 0.44 and 0.88 BLEU points for the Java and Python datasets respectively.

\smallskip\noindent
\textbf{Impact of position representation.}
\label{subsec:impact_of_pos_rep}
We perform an ablation study to investigate the benefits of encoding the absolute position of code tokens or modeling their pairwise relationship for the source code summarization task, and the results are presented in Table \ref{table:abs_position_ablation} and \ref{table:rel_position_ablation}.
Table \ref{table:abs_position_ablation} demonstrates that learning the absolute position of code tokens are not effective as we can see it slightly hurts the performance compared to when it is excluded. 
This empirical finding corroborates the design choice of \citet{iyer2016summarizing}, where they did not use the sequence information of the source code tokens.

\input{table/table4}

On the other hand, we observe that learning the pairwise relationship between source code tokens via relative position representations helps as Table \ref{table:rel_position_ablation} demonstrates higher performance. 
We vary the clipping distance, $k$, and consider ignoring the directional information while modeling the pairwise relationship.
The empirical results suggest that the directional information is indeed important while 16, 32, and $2^i$ relative distances result in similar performance (in both experimental datasets).

\input{table/table5}

\smallskip\noindent
\textbf{Varying model size and number of layers.}
We perform ablation study by varying $d_{model}$ and $l$ and the results are presented in Table \ref{table:hid_and_layer_ablation}.\footnote{Considering the model complexity, we do not increase the model size or number of layers further.}
In our experiments, we observe that a deeper model (more layers) performs better than a wider model (larger $d_{model}$).
Intuitively, the source code summarization task depends on more semantic information than syntactic, and thus deeper model helps.

\smallskip\noindent
\textbf{Use of Abstract Syntax Tree (AST).}
We perform additional experiments to employ the abstract syntax tree (AST) structure of source code in the Transformer.
We follow \citet{hu2018deep} and use the Structure-based Traversal (SBT) technique to transform the AST structure into a linear sequence. 
We keep our proposed Transformer architecture intact, except in the copy attention mechanism, we use a mask to block copying the non-terminal tokens from the input sequence.
It is important to note that, with and without AST, the average length of the input code sequences is 172 and 120, respectively. 
Since the complexity of the Transformer is $O(n^2 \times d)$ where $n$ is the input sequence length, hence, the use of AST comes with an additional cost.
Our experimental findings suggest that the incorporation of AST information in the Transformer does not result in an improvement in source code summarization.
We hypothesize that the exploitation of the code structure information in summarization has limited advantage, and it diminishes as the Transformer learns it implicitly with relative position representation.


\input{table/qual_ex}

\smallskip\noindent
\textbf{Qualitative analysis.}
We provide a couple of examples in Table \ref{table:qual_examples} to demonstrate the usefulness of our proposed approach qualitatively (more examples are provided in Table \ref{app:java_ablation_example} and \ref{app:python_ablation_example} in the Appendix).
The qualitative analysis reveals that, in comparison to the Vanilla Transformer model, the copy enabled model generates shorter summaries with more accurate keywords. 
Besides, we observe that in a copy enabled model, frequent tokens in the code snippet get a higher copy probability when relative position representations are used, in comparison to absolute position representations.
We suspect this is due to the flexibility of learning the relation between code tokens without relying on their absolute position.

%% file: table/table4.tex
\begin{table}[t]
\centering
\resizebox{\linewidth}{!}{%
\begin{tabular}{c c|c c c}
\hline
Source & Target & BLEU & METEOR & ROUGE-L \\ 
\hline
\cmark & \cmark & 43.41 & 25.91 & 52.71  \\
\cmark & \xmark & 42.34 & 24.74 & 50.96 \\
\xmark & \cmark & {\bf 43.59} & {\bf 26.00} & {\bf 52.88} \\
\xmark & \xmark & 41.85 & 24.32 & 50.87 \\
\hline
\end{tabular}
}
\caption{Ablation study on absolute positional representations using the ``Base Model'' on the Java dataset.}
\label{table:abs_position_ablation}
\vspace{2mm}
\end{table}

\begin{table}[t]
\centering
\resizebox{\linewidth}{!}{%
\begin{tabular}{c c|c c c}
\hline
$k$ & Directional & BLEU & METEOR & ROUGE-L \\ 
\hline
\multirow{2}{*}{8} & \cmark & 44.22 & 26.35 & 53.86  \\
& \xmark & 42.61 & 24.67 & 51.10 \\
\hline
\multirow{2}{*}{16} & \cmark & 44.14 & 26.34 & 53.95 \\
& \xmark & 44.06 & 26.31 & 53.51 \\
\hline
\multirow{2}{*}{32} & \cmark & {\bf 44.55} & {\bf 26.66} & {\bf 54.30} \\
& \xmark & 43.95 & 26.28 & 53.24 \\
\hline
\multirow{2}{*}{$2^i$} & \cmark & 44.37 & 26.58 & 53.96 \\
& \xmark & 43.58 & 25.95 & 52.73 \\
\hline
\end{tabular}
}
\caption{Ablation study on relative positional representations (in encoding) for Transformer.
While 8, 16, and 32 represents a fixed relative distance for all the layers, $2^i$ (where $i = 1, \ldots, L$; $L=6$) represents a layer-wise relative distance for Transformer.
}
\label{table:rel_position_ablation}
\end{table}

%% file: table/table5.tex
\begin{table}[!ht]
\centering
\resizebox{\linewidth}{!}{%
\begin{tabular}{c|c|c c c}
\hline
 & \#Param. & BLEU & METEOR & ROUGE-L \\ 
\hline
\multicolumn{5}{l}{Varying the model size ($d_{model}$)} \\
\hline
256 & 15.8 & 38.21 & 21.54 & 48.63 \\
384 & 28.4 & 41.71 & 24.51 & 51.42 \\
512 & 44.1 & 43.41 & 25.91 & 52.71 \\
768 & 85.1 & {\bf 45.29} & {\bf 27.56} & {\bf 54.39} \\
\hline
\multicolumn{5}{l}{Varying the number of layers ($l$)} \\
\hline
3 & 22.1 & 41.26 & 23.54 & 51.37 \\
6 & 44.1 & 43.41 & 25.91 & 52.71 \\
9 & 66.2 & 45.03 & 27.21 & 54.02 \\
12 & 88.3 & {\bf 45.56} & {\bf 27.64} & {\bf 54.89} \\
\hline
\end{tabular}
}
\caption{Ablation study on the hidden size and number of layers for the ``Base Model'' on the Java dataset. 
We use $d_{model} = H$, $d_{ff} = 4H$, $h = 8$, and $d_k = d_v = 64$ in all settings. 
We set $l = 6$ and $d_{model} = 512$ while varying $d_{model}$ and $l$ respectively.
\#Param. represents the number of trainable parameters in millions (only includes Transformer parameters).
}
\label{table:hid_and_layer_ablation}
\end{table}

%% file: table/qual_ex.tex
\begin{table*}[!ht]
\centering
\small
\begin{tabular}{ p{0.95\linewidth}}
\hline

\vspace{-5mm}
\begin{minted}[fontsize=\footnotesize
% ,xleftmargin=17pt,linenos
]{java}
public static String selectText(XPathExpression expr, Node context) { 
    try { 
        return (String)expr.evaluate(context, XPathConstants.STRING ); 
    } catch (XPathExpressionException e) { 
        throw new XmlException(e); 
    } 
}
\end{minted}
\vspace{-1mm}
{\color{blue} Base Model}: evaluates the xpath expression to a xpath expression .\\
\vspace{-2mm}
{\color{purple} Full Model w/o Relative Position}: evaluates the xpath expression .\\
\vspace{-2mm}
{\color{orange} Full Model w/o Copy Attention Attention}: evaluates the xpath expression as a single element .\\
\vspace{-2mm}
{\color{cyan} Full Model:} evaluates the xpath expression as a text string .\\
\vspace{-2mm}
{\color{violet} Human Written:} evaluates the xpath expression as text .\\
\hline

\vspace{-5mm}
\begin{minted}[fontsize=\footnotesize
% ,xleftmargin=17pt,linenos
]{python}
def get_hosting_service(name):
    try:
        return hosting_service_registry.get(u'hosting service id', name)
    except ItemLookupError:
        return None
\end{minted}
\vspace{-1mm}
{\color{blue} Base Model}: returns the color limits from the current service name .\\
\vspace{-2mm}
{\color{purple} Full Model w/o Relative Position}: return the hosting service .\\
\vspace{-2mm}
{\color{orange} Full Model w/o Copy Attention}: return the name of the service .\\
\vspace{-2mm}
{\color{cyan} Full Model :} return the hosting service name .\\
\vspace{-2mm}
{\color{violet} Human Written:} return the hosting service with the given name .\\
\hline

\end{tabular}
\caption{Qualitative example of different models' performance on Java and Python datasets.}
\label{table:qual_examples}
\end{table*}

%% file: relwork.tex
\section{Related Work}
Most of the neural source code summarization approaches frame the problem as a sequence generation task and use recurrent encoder-decoder networks with attention mechanisms as the fundamental building blocks \cite{iyer2016summarizing, liang2018automatic, hu2018deep, hu2018summarizing}.
Different from these works, \citet{allamanis2016convolutional} proposed a convolutional attention model to summarize the source codes into short, name-like summaries.

Recent works in code summarization utilize structural information of a program in the form of \emph{Abstract Syntax Tree} (AST) that can be encoded using tree structure encoders such as Tree-LSTM \cite{shido2019automatic}, Tree-Transformer \cite{harer2019tree}, 
and Graph Neural Network \cite{leclair2020improved}.
In contrast, \citet{hu2018deep} proposed a structure based traversal (SBT) method to flatten the AST into a sequence and showed improvement over the AST based methods.
Later, \citet{leclair2019subroutines} used the SBT method and decoupled the code structure from the code tokens to learn better structure representation.

Among other noteworthy works, API usage information \cite{hu2018summarizing}, reinforcement learning \cite{wan2018improving}, dual learning \cite{wei2019code}, retrieval-based techniques \cite{zhangretrieval} are leveraged to further enhance the code summarization models.
We can enhance a Transformer with previously proposed techniques; however, in this work, we limit ourselves to study different design choices for a Transformer without breaking its' core architectural design philosophy.

%% file: conclusion.tex
\section{Conclusion}
This paper empirically investigates the advantage of using the Transformer model for the source code summarization task.
We demonstrate that the Transformer with relative position representations and copy attention outperforms state-of-the-art approaches by a large margin.
In our future work, we want to study the effective incorporation of code structure into the Transformer and apply the techniques in other software engineering sequence generation tasks (e.g., commit message generation for source code changes).

%% file: appendix.tex









\appendix

\addcontentsline{toc}{section}{Appendices}
\renewcommand{\thesubsection}{\Alph{subsection}}

\subsection{Hyper-Parameters}
\label{app:hp}

Table \ref{table:hyperparameters} summarizes the hyper-parameters that we used in our experiments.

\input{table/table6}

\subsection{Recurrent Encoder-Decoder vs. Transformer on Python Dataset}
\label{app:rnn_vs_trans}

\begin{table}[!ht]
\centering
\resizebox{\linewidth}{!}{%
\small
\begin{tabular}{l|c@{\hskip 0.1in} c@{\hskip 0.1in} cc@{\hskip 0.1in}}
\hline
Models & BLEU & METEOR & ROUGE-L \\ 
\hline
Seq2seq & 30.57 & 17.86 & 43.64 \\
Seq2seq$^\ast$ & 29.08 & 17.12 & 42.97 \\
Transformer & 31.08 & 18.57 & 44.31 \\
Transformer$^\ast$ & {\bf 31.38} & {\bf 18.69} & {\bf 44.68} \\
\hline
\end{tabular}
}
\caption{
Comparison between recurrent sequence-to-sequence (Seq2seq) model and Transformer on the Python dataset.
$^\ast$ indicates models are equipped with the copy attention mechanism.
}
\label{table:rnn_vs_trans}
\end{table}

While conducting our study using the Transformer on the Python dataset, we observed a significant gain over the state-of-the-art methods as reported in \citet{wei2019code}.
However, our initial experiments on this dataset using recurrent sequence-to-sequence models also demonstrated higher performance compared to the results report in \citet{wei2019code}.
We suspect that such lower performance is due to not tuning the hyper-parameters correctly.
So for the sake of fairness and to investigate  the true advantages of Transformer, we present a comparison on recurrent Seq2seq model and Transformer in Table \ref{table:rnn_vs_trans} using our implementation.\footnote{Our implementation is based on Open-NMT \cite{klein2017opennmt} and PyTorch 1.3.}

We can see from Table \ref{table:rnn_vs_trans}, the performance of the recurrent Seq2seq model is much better than the results reported in prior works.
However, to our surprise, the copy attention mechanism does not result in improvement for the recurrent Seq2seq model. When we looked into the training perplexity and the validation performance, we also observed lower performance in comparison to the base recurrent Seq2seq model.
In comparison, our proposed Transformer-based approach outperforms the recurrent Seq2seq models by a large margin showing its effectiveness for source code summarization.


\onecolumn

\subsection{Qualitative Examples}
\label{app:qual_exam}

\input{table/java_ablation}
\input{table/python_ablation}


%% file: table/table6.tex
\begin{table}[!ht]
\centering
\begin{tabular}{c | c | c}
\hline
 & Hyper-parameter & Value \\ 
\hline
\multirow{1}{*}{Embedding}  & $k$ & 16 \\
\hline
\multirow{4}{*}{Model}  & $l$ & 6 \\
& $h$ & 8 \\
& $d_{model}$ & 512 \\
& $d_k, d_v$ & 64 \\
& $d_{ff}$  & 2048 \\
\hline 
\multirow{4}{*}{Training}  & dropout & 0.2 \\
& optimizer & Adam \\
& learning rate  & 0.0001 \\
& batch size  & 32 \\
\hline
\multirow{1}{*}{Testing}  & beam size & 4 \\
\hline 
\end{tabular}
\caption{Hyper-parameters in our experiments. 
$l$ and $h$ indicates the number of layers and heads in Transformer respectively.
$k$ refers to the clipping distance in relative position representations in Transformer.
}
\label{table:hyperparameters}
\end{table}

%% file: table/java_ablation.tex
\begin{table}[H]
\centering
\small
\begin{tabular}{ p{0.95\linewidth}}
\hline

\vspace{-5mm}
\begin{minted}[]{java}
public static terminal find(String with_name) { 
    if(with_name == null) 
        return null; 
    else 
        return (terminal)all.get(with_name); 
}
\end{minted}
\vspace{-2mm}
{\color{blue} Base Model}: lookup a non terminal by name string\\
\vspace{-2mm}
{\color{purple} Full Model w/o Relative Position}: lookup a terminal terminal by name string\\
\vspace{-2mm}
{\color{orange} Full Model w/o Copy Attention}: lookup a non terminal by name string\\
\vspace{-2mm}
{\color{cyan} Full Model:} lookup a terminal by name\\
\vspace{-2mm}
{\color{violet} Human Written:} lookup a terminal by name string .\\
\hline

\vspace{-5mm}
\begin{minted}[]{java}
public static String selectText(XPathExpression expr, Node context) { 
    try { 
        return (String)expr.evaluate(context, XPathConstants.STRING ); 
    } catch (XPathExpressionException e) { 
        throw new XmlException(e); 
    } 
}
\end{minted}
\vspace{-2mm}
{\color{blue} Base Model}: evaluates the xpath expression to a xpath expression .\\
\vspace{-2mm}
{\color{purple} Full Model w/o Relative Position}: evaluates the xpath expression .\\
\vspace{-2mm}
{\color{orange} Full Model w/o Copy Attention Attention}: evaluates the xpath expression as a single element .\\
\vspace{-2mm}
{\color{cyan} Full Model:} evaluates the xpath expression as a text string .\\
\vspace{-2mm}
{\color{violet} Human Written:} evaluates the xpath expression as text .\\
\hline

\vspace{-5mm}
\begin{minted}[]{java}
public CTaggingPanel(
    final JFrame parent, final ZyGraph graph, final ITagManager manager) { 
    super(new BorderLayout()); 
    mtagsTree = new CTagsTree(parent, graph, manager); 
    final JScrollPane pane = new JScrollPane(mtagsTree); 
    pane.setVerticalScrollBarPolicy(
        ScrollPaneConstants.VERTICAL_SCROLLBAR_AS_NEEDED); 
    pane.setHorizontalScrollBarPolicy(
        ScrollPaneConstants.HORIZONTAL_SCROLLBAR_AS_NEEDED); 
    add(pane); 
    setBorder(new TitledBorder(new LineBorder(Color.LIGHT_GRAY, NUM, BOOL), STRING)); 
    setDoubleBuffered(BOOL); 
}
\end{minted}
\vspace{-2mm}
{\color{blue} Base Model}: creates a new dnetscapesslservername dialog .\\
\vspace{-2mm}
{\color{purple} Full Model w/o Relative Position}: creates a new settings dialog .\\
\vspace{-2mm}
{\color{orange} Full Model w/o Copy Attention}: creates a new toolbar panel .\\
\vspace{-2mm}
{\color{cyan} Full Model:} creates a new api panel object .\\
\vspace{-2mm}
{\color{violet} Human Written:} creates a new panel object .\\
\hline

\vspace{-5mm}
\begin{minted}[]{java}
public DSignCsr(JFrameparent, PKCS10CertificationRequest pkcs10Csr, File csrFile, 
                PrivateKey signPrivateKey, KeyPairType signKeyPairType, 
                X509Certificate verificationCertificate, Provider provider) 
                throws CryptoException{ 
    super(parent, Dialog.ModalityType.DOCUMENT_MODAL); 
    this.pkcs10Csr = pkcs10Csr; 
    this.csrFile = csrFile; 
    this.signPrivateKey = signPrivateKey; 
    this.signKeyPairType = signKeyPairType; 
    this.verificationCertificate = verificationCertificate; 
    this.provider = provider; 
    setTitle(res.getString(STRING)); 
    initComponents(); 
}
\end{minted}
\vspace{-2mm}
{\color{blue} Base Model}: creates a new dsigncsr dialog for a spkac formatted csr .\\
\vspace{-2mm}
{\color{purple} Full Model w/o Relative Position}: creates a new signer dialog for a pkcs \# 10 formatted .\\
\vspace{-2mm}
{\color{orange} Full Model w/o Copy Attention}: creates a new dsigncsr dialog for a spkac formatted csr .\\
\vspace{-2mm}
{\color{cyan} Full Model:} creates a new dsigncsr dialog for a pkcs \# 10 formatted csr .\\
\vspace{-2mm}
{\color{violet} Human Written:} creates a new dsigncsr dialog for a pkcs \# 10 formatted csr .\\
\hline

\end{tabular}
\caption{Qualitative example of different models' performance in Java dataset.}
\label{app:java_ablation_example}
\end{table}

%% file: table/python_ablation.tex
\begin{table}[H]
\centering
\small
\begin{tabular}{ p{0.95\linewidth}}
\hline

\vspace{-5mm}
\begin{minted}[]{python}
def get_hosting_service(name):
    try:
        return hosting_service_registry.get(u'hosting service id', name)
    except ItemLookupError:
        return None
\end{minted}
\vspace{-2mm}
{\color{blue} Base Model}: returns the color limits from the current service name .\\
\vspace{-2mm}
{\color{purple} Full Model w/o Relative Position}: return the hosting service .\\
\vspace{-2mm}
{\color{orange} Full Model w/o Copy Attention}: return the name of the service .\\
\vspace{-2mm}
{\color{cyan} Full Model :} return the hosting service name .\\
\vspace{-2mm}
{\color{violet} Human Written:} return the hosting service with the given name .\\
\hline

\vspace{-5mm}
\begin{minted}[]{python}
def save_pickle(obj, fname):
    with get_file_obj(fname, 'wb') as fout: 
        cPickle.dump(obj, fout, protocol=-1)
\end{minted}
\vspace{-2mm}
{\color{blue} Base Model}: pickle object obj to file fname .\\
\vspace{-2mm}
{\color{purple} Full Model w/o Relative Position}: save object to file .\\
\vspace{-2mm}
{\color{orange} Full Model w/o Copy Attention}: raw\_data: object obj to file fname .\\
\vspace{-2mm}
{\color{cyan} Full Model :} save object to file fname .\\
\vspace{-2mm}
{\color{violet} Human Written:} save the object to file via pickling .\\
\hline

\vspace{-5mm}
\begin{minted}[]{python}
def get_temp_dir:
    temp = get_environ_variable('TMP') 
    if temp is None: 
        temp = get_environ_variable('TEMP') 
    if temp is None or '' in temp and os.name == 'nt':
        temp = 'C \\temp'
    if temp None or '' in temp and os.name == 'posix':
        temp = '/tmp'
    return temp
\end{minted}
\vspace{-2mm}
{\color{blue} Base Model}: returns the name of the sample environment variable .\\
\vspace{-2mm}
{\color{purple} Full Model w/o Relative Position}: returns the next temporary directory of a file .\\
\vspace{-2mm}
{\color{orange} Full Model w/o Copy Attention}: get the directory related to store the stubbed .\\
\vspace{-2mm}
{\color{cyan} Full Model :} return a temporary filename .\\
\vspace{-2mm}
{\color{violet} Human Written:} returns a temporary directory .\\
\hline

\vspace{-5mm}
\begin{minted}[]{python}
def get_exploration_memcache_key(exploration_id, version=None):
    if version:
        return 'exploration-version %s %s' % exploration_id, version 
    else:
        return 'exploration %s' % exploration_id
\end{minted}
\vspace{-2mm}
{\color{blue} Base Model}: returns the key for an instance for the project .\\
\vspace{-2mm}
{\color{purple} Full Model w/o Relative Position}: returns a memcache key for the given version .\\
\vspace{-2mm}
{\color{orange} Full Model w/o Copy Attention}: returns a memcache for the exploration id .\\
\vspace{-2mm}
{\color{cyan} Full Model :} returns a memcache key for the specified exploration .\\
\vspace{-2mm}
{\color{violet} Human Written:} returns a memcache key for an exploration .\\
\hline

\vspace{-5mm}
\begin{minted}[]{python}
def get_svc_avail_path(): 
    return AVAIL_SVR_DIRS
\end{minted}
\vspace{-2mm}
{\color{blue} Base Model}: get the actual path .\\
\vspace{-2mm}
{\color{purple} Full Model w/o Relative Position}: returns a list of services .\\
\vspace{-2mm}
{\color{orange} Full Model w/o Copy Attention}: return a list of services that are available .\\
\vspace{-2mm}
{\color{cyan} Full Model :} returns a list of available services .\\
\vspace{-2mm}
{\color{violet} Human Written:} return list of paths that may contain available services .\\
\hline

\vspace{-5mm}
\begin{minted}[]{python}
def volume_attach(provider, names, **kwargs):
    client.get_client_info() 
    client.extra_action(provider=provider, names=names, action='volume attach', 
                        **kwargs) 
    return info
\end{minted}
\vspace{-2mm}
{\color{blue} Base Model}: attempt to attach volume .\\
\vspace{-2mm}
{\color{purple} Full Model w/o Relative Position}: attach volume cli example: .\\
\vspace{-2mm}
{\color{orange} Full Model w/o Copy Attention}: attach volume cli example: .\\
\vspace{-2mm}
{\color{cyan} Full Model :} attach volume information cli example: .\\
\vspace{-2mm}
{\color{violet} Human Written:} attach volume to a server cli example: .\\
\hline

\end{tabular}
\caption{Qualitative example of different models' performance in Python dataset.}
\label{app:python_ablation_example}
\end{table}